# Solution-Grown Nanowire Devices for Sensitive and Fast Photo Detection

*Alexander Littig, Hauke Lehmann, Christian Klinke, Tobias Kipp\*, and Alf Mews*

Institute of Physical Chemistry, University of Hamburg,

Grindelallee 117, 20146 Hamburg, Germany

E-mail: kipp@chemie.uni-hamburg.de

ABSTRACT

Highly sensitive and fast photodetector devices with CdSe quantum nanowires as active elements have been developed exploiting the advantages of electro- and wet-chemical routes. Bismuth nanoparticles electrochemically synthesized directly onto interdigitating platinum electrodes serve as catalysts in the following solution-liquid-solid synthesis of quantum nanowires directly on immersed substrates under mild conditions at low temperature. This fast and simple preparation process leads to a photodetector device with a film of nanowires of limited thickness bridging the electrode gaps, in which a high fraction of individual nanowires are electrically contacted and can be exposed to light at the same time. The high sensitivity of the photodetector device can be expressed by its on/off-ratio or its photosensitivity of more than $10^7$ over a broad wavelength range up to about 700 nm. The specific detectivity and responsivity are determined to $D^* = 4 \cdot 10^{13}$ Jones and $R = 0.32$ A/W, respectively. The speed of the device reflects itself in a 3 dB frequency above 1 MHz corresponding to rise and fall times below 350 ns. The remarkable combination of a high sensitivity and a fast response is attributed to depletion regions inside the nanowires, tunnel-junction barriers between nanowires, as well as Schottky contacts at the electrodes, where all these features are strongly influenced by the number of photo generated charge carriers.

INTRODUCTION

Due to their distinguished performance in photonics, electronics, and optoelectronics, semiconductor nanowires (NWs) are promising candidates as active components and building blocks for nanoscale devices. Hence, over the last two decades several bottom-up fabrication routes for different types of NWs with precisely controlled morphologies have been developed. With those NWs, novel nanodevices such as gas sensors,[1–5] photodetectors,[6–9] photovoltaic devices,[10,11] light-emitting diodes,[12,13] image-sensor circuits,[14] lasers,[15] field-effect transistors,[16,17] thermoelectric devices,[18] or logic gates[19] were developed. Most of these devices rely on rather thick NWs fabricated by gas-phase techniques, with diameters in the range of several tens of nanometers. As an alternative, very thin NWs with diameters in the range of the bulk exciton dimension can be highly advantageous for device operation since they exhibit a higher surface-to-volume ratio and quantization effects to tailor the device properties.[20] Especially for II-IV semiconductors such as CdSe it has been shown that such thin and highly crystalline nanowires can be simply synthesized by wet chemistry following the solution-liquid-solid (SLS) approach.[21] These nanowires exhibit high crystallinity, as has been shown in high resolution transmission electron microscopy (TEM) measurements.[20,22-26] By variation of the reaction conditions, their diameter can be easily tuned to values below the exciton dimension (i.e. < 12 nm for CdSe).[27] Hence, they show optical properties similar to quantized semiconductor nanoparticles, such as pronounced blue shift of the absorption and emission, and fluorescence blinking. Thus, they are called quantum nanowires (QNWs).[26] Devices assembled with such QNWs as building blocks combine the advantages of a directional electrical transport along the macroscopic direction of the NW under quantum confinement with a very high surface-to-volume ratio, which makes them highly sensitive to environmental conditions and thus suitable for sensor applications.

In general, the fabrication of a NW device is a multistep process.[28] In most cases, the procedure starts with the synthesis of the specific NWs after which several steps of purification have to be performed before they can be dispersed on substrates. Then, lithography steps are necessary to contact the wires. Thus, this procedure is quite expensive and time-consuming and a challenge for mass production. Hence, several on-chip growth methods for NWs have been developed, which enable fabrication processes of those devices by fewer steps. This was demonstrated for $SnO_2$ and $CdS_xSe_{1-x}$ NWs based on the vapor-liquid-solid (VLS) synthesis.[3,29] While these methods clearly reduce the costs and efforts for device preparation, still high reaction temperatures are required, which limit the variety of suitable substrates. In addition, NWs grown by the VLS method are typically thicker than NWs grown in solution by the SLS method.[30] Solution-based synthesis methods offer an easy way to horizontally grow QNWs at temperatures between 180 and 300 °C on substrates like silicon, glass, or flexible plastic.[25,31] Like for the VLS method, also for the SLS method, liquid metal droplets act as catalysts during NW growth. Here, due to the low reaction temperatures, a metal with a low melting point, typically bismuth, is used. For a growth of NWs directly on the electrodes, the bismuth can be deposited on the electrode surface by different methods like (i) the thermal evaporation of a thin Bi film, (ii) spin-coating of colloidal Bi nanoparticles (NPs), or (iii) the electrochemical preparation of Bi NPs. In the first method, the bismuth film melts in the solution before or during the growth reaction and forms active bismuth droplets.[22] As a drawback, the droplets have a broad size distribution which can lead to a corresponding broad diameter distribution of the grown NWs. In the second method, the narrow size distribution of colloidal Bi NPs is exploited,[24] which can lead to NWs with a very narrow diameter distribution. However, it remains challenging to deposit the NPs only on the predefined electrode structures and also to enable a good electronic contact between electrodes and NWs. The third method elegantly circumvents above problems. Here, Bi NPs are synthesized directly and exclusively on the electrodes without any further fabrication or purification steps by electrochemical deposition. The size and the density of the Bi NPs can be tuned by parameters like reduction potential, reduction time or concentration of the ionic bismuth precursor solution.[32] Finally the electrochemical reduction of bismuth at the electrode leads to a direct contact between the Bi NPs and the conductive substrate, which reduces the probability of dissolution from the surface during the synthesis and enables a low contact resistance between the NW and the electrode.

In the present work, we report on QNWs synthesized directly on electrodes and employing them as active elements in a photodetector and in an optical switch. To achieve this, we deposited Bi NPs electrochemically on an interdigitating platinum electrode structure with widths and gaps of 10 μm, respectively. The photosensitive electrode structure covering an area of 1 mm² was fabricated on a glass substrate by optical lithography using standard photoresists and sputtering of tungsten as adhesion layer and platinum as top material. Under mild conditions, i.e. at a temperature of 200 °C, thin CdSe QNWs were synthesized horizontally on the electrode by employing the SLS method. By using this simple and fast fabrication route we built a high performance device which acts as a photodetector with small dark currents, high on/off-ratios, high specific detectivity as well as a high bandwidth with corresponding fast rise and fall times.

RESULTS AND DISCUSSION

**Preparation and Characterization of CdSe QNWs.** In order to prepare a photodetector device, we firstly deposited Bi particles electrochemically on the electrodes and subsequently we grew CdSe QNWs from the substrates immersed in solution. In a prior work[32] we demonstrated that the size and density of the Bi particles depend on several reaction parameters such as electrochemical reduction potential and deposition time, on the chemical nature and the concentrations of the precursors used, and finally also on the electrode material and its surface roughness. Here, we deposited the Bi NPs electrochemically on the Pt electrodes by using a reduction potential of -300 mV for 100 s with a three electrode setting from a 1 mM aqueous bismuth solution (details can be found in the experimental section). Figure 1a shows an atomic-force microscope (AFM) image of the dense film of Bi catalyst NPs that was required to prepare a thick and dense network of QNWs. Even though the high density of the particles does not allow for a detailed determination of the size distribution by AFM, it can be estimated that a typical size of the bigger particles is of the order of 15-30 nm. Using these Bi NPs as catalysts, CdSe QNWs were synthesized at a growth temperature of 200 °C for 20 min following the procedure described in Ref. 27.

Figure 1b and c show scanning–electron microscope (SEM) images of the prepared QNW networks. The micrograph in panel b shows an area on top of the electrode with the same magnification as the AFM image in a. In the lower-magnification image in panel c, electrode and gap areas can be distinguished, both homogeneously covered with an equally dense QNW film. It is an interesting intrinsic feature of our devices that the film height is restricted by the self-limited growth process which depends on the amount of bismuth catalyst. As a consequence virtually all QNWs have their origin in Bi catalyst particles with a direct contact to the electrode.

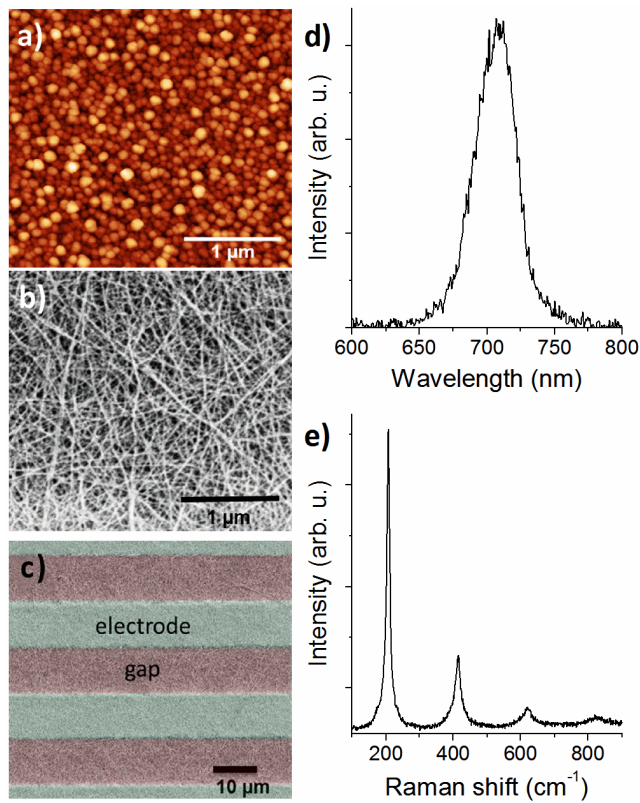

Figure 1. (a) AFM image of bismuth NPs on top of a platinum electrode. (b) SEM image of CdSe QNWs on the electrode. (c) SEM overview of QNWs on the interdigitating electrode structure. As a guide for the eyes, electrodes and gaps are colored greenish and reddish, respectively. (d) PL spectrum of the QNWs. (e) Raman spectrum of the QNWs.

Photoluminescence (PL) spectra of these CdSe QNWs exhibit a peak maximum wavelength of 706 nm, as is exemplarily shown in Figure 1d. It is slightly blue-shifted compared to the bulk CdSe emission at 713 nm caused by the radial quantum confinement of the photo-generated excitons. Through a correlation of QNW diameter and emission wavelength, we estimate the mean diameter of the QNWs to be in the range of 12 nm or below.[20] The diameter can directly and more accurately be determined by TEM, however therefor the device has to be destroyed. Apart from the device on which all the measurements presented here have been obtained, a second device has been fabricated, using the same synthesis parameters. We detached the nanowires from this sample by ultrasound treatment and determined the diameter distribution of the QNW fragments by TEM to be 9.1 ± 1.9 nm. Both measurements, i.e., PL on the original sample and TEM on the comparison sample, reveal diameters of the nanowires within the quantum-confinement regime. In general, the diameter of the SLS grown NWs is strongly dependent on the catalyst NP sizes.[30,33–35] The difference between the QNW diameter and the diameter of the larger Bi NPs of up to 30 nm suggests that only the smaller Bi NPs or only domains of NPs are catalytically active during the QNW synthesis. This can be explained by the decreasing melting temperature for Bi NPs with decreasing size, down to 150 °C for approximately 4 nm particles.[36,37]

Figure 1e exemplarily shows a Raman spectrum of the CdSe QNWs. It exhibits the characteristic LO phonon mode at 208 cm$^{-1}$ and its overtones (2LO) at 414 cm$^{-1}$, (3LO) at 620 cm$^{-1}$, and (4LO) at 826 cm$^{-1}$.[38] The photoluminescence as well as the Raman spectra verify the successful preparation of high quality CdSe QNWs.

**CdSe-QNW Device as a Photodetector.** Our QNW device acts as a visible-light photodetector. The spectral response is shown in the photocurrent spectra in Figure 2a. Obviously, the device's sensitivity is fairly stable over a broad range of wavelengths between 500 and 675 nm. The spectrum exhibits a photocurrent drop in the subsequent wavelength range of 675 to 710 nm, corresponding to the NWs optical band gap, as supported by the photoluminescence emission wavelength of 706 nm. The small current at lower energies than the bandgap energy might be due to defect states.

For comparability all following photocurrent measurements have been performed by using a excitation laser with a wavelength of 637 nm.

Figure 2b shows the I-V characteristics of the CdSe QNW device for darkness (lowest curve) and for increasing light intensities. The induced photocurrent, measured under ambient conditions, strongly depends on the incident light. These I-V curves show a symmetric photocurrent behavior for positive and negative voltages, since both electrodes are made of platinum. Note that the slight horizontal shift of the dark-current curve is an artifact induced by the extremely low current.

Figure 2c correlates the measured photocurrent for an applied source-drain voltage of 10 V to the used excitation laser power density on a double logarithmic scale. The grey area represents the dark-current level. While the dark current is in the range of 1-3 pA the photocurrent reaches values of more than 30 µA under an incident laser power density of 120 mW/cm$^2$. Over the whole measured light power densitiy range of about six orders of magnitude, the relation between photocurrent ($I_p$) and laser power density ($P$) follows a simple power-law behavior:

$$I_p \propto P^x \qquad (1)$$

with the parameter $x = 0.8$, as determined by fitting the slope in Figure 2c. Such power laws have been observed for different kinds of photoconductors, such as thin CdSe films based on quantum dots and CdSe quantum NW films, but with lower on/off-ratios.[8,39] They have also been found for NW devices built-up from single or just a few NWs, but only for a limited range of light intensities.[17,31,40,41] While a simple linear dependency of $I_p$ on $P$ would result in $x = 1$, a power-law exponent smaller than unity, which is often observed, is discussed to point to complex electron-hole generation, trapping, and recombination processes within the semiconductor.[40]

Very recently, for NW devices based on ZnO and PbS, Ullrich et al. reported on a change of the power-law exponent from close to unity for smaller illumination intensities to 0.5 for very high illumination powers.[42] Interestingly, the power-law exponent of 0.8 for our QNWs devices did not changed over the whole range of illumination intensities. However, we cannot exclude any change in the exponent for even larger illumination intensities than the maximum of 240 mW/cm$^2$ as used in our experiments.

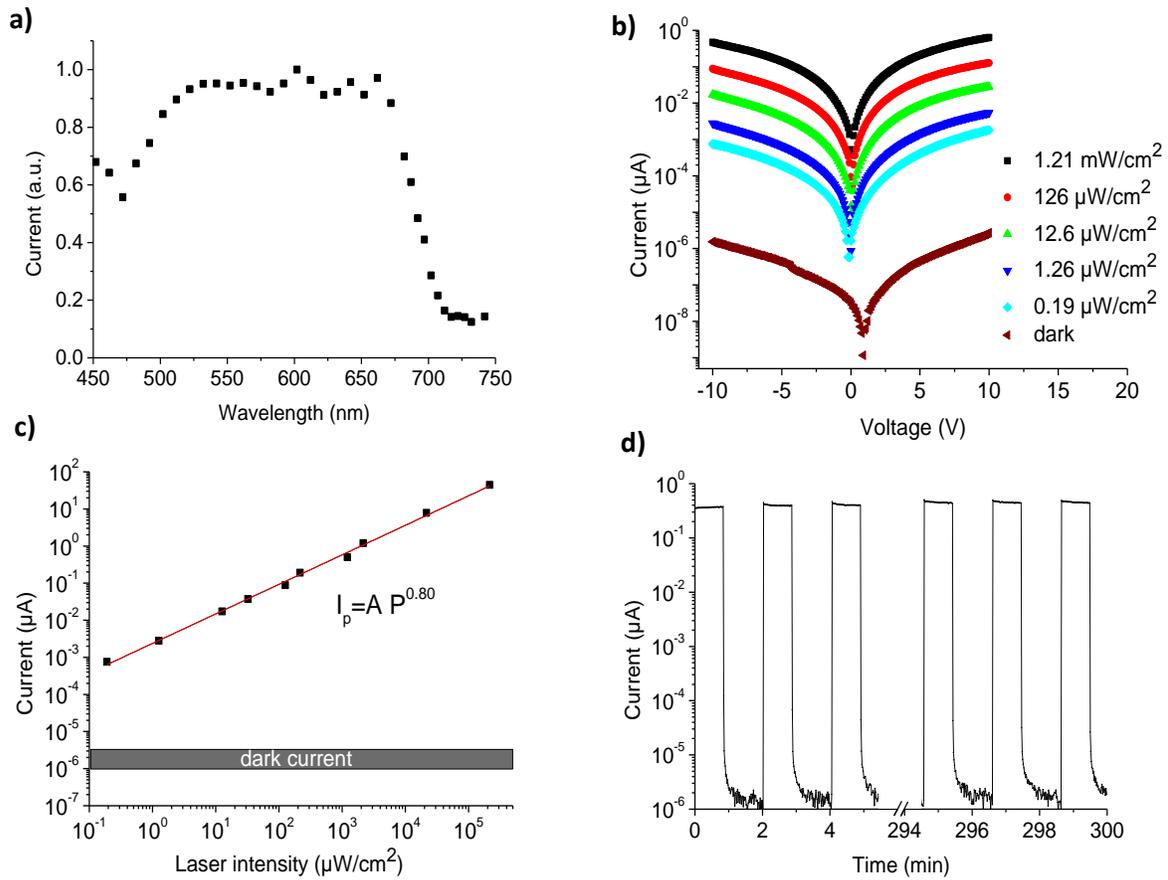

Figure 2 Photocurrent measurements under ambient conditions. (a) Photocurrent spectra recorded with an applied source-drain voltage of 10 V. The drop of the photocurrent at 700 nm is at a similar wavelength as the PL emission wavelength. (b) I-V curves of the CdSe-QNW device at different light power densities ($\lambda$ = 637 nm) on a logarithmic scale. The curve of the dark current is slightly asymmetric, due to the extreme low current below 100 fA in the voltage region from -2 to 2 V. (c) Variation of the photocurrent for different light power densities between 0.19 $\mu W/cm^2$ and 240 $mW/cm^2$ with an applied source-drain voltage of 10 V. The exponent of the power-law fit was determined to 0.8. The gray area symbolizes the dark-current level. (d) Long-term stability measurement with an incident laser power density of 0.65 $mW/cm^2$. The laser was electrically chopped with an asymmetric frequency of about 8 mHz. The source-drain voltage was set to 10 V.

Figure 2c also allows for a quantification of the laser-power dependent sensitivity of our photocurrent device. A frequently used measure of the sensitivity is the on/off-ratio. For example, at a light power density of 120 mW/cm² the on/off-ratio exceeds $10^7$. To the best of our knowledge, this is the highest on/off-ratio for NW networks or NW films for visible light so far. Another commonly used figure of merit for the sensitivity is the so called photosensitivity which is the quotient $S = (I_{photo} - I_{dark}) / I_{dark}$. Since, in our case $I_{dark}$ is extremely small (< 3 pA) the photosensitivity directly equals our on/off-ratio.

Another characteristic variable of a photodetector is the spectral responsivity, defined as:

$$R_\lambda = \frac{I_p}{P \cdot A} = EQE \frac{e}{h\upsilon} \qquad (2)$$

where *A* is the area, *EQE* is the external quantum efficiency, *P* is again the light power density, *e* is the electron charge, *h* is the Planck's constant and $\upsilon$ is the frequency of the incident light. Note that *EQE* is equal to the so-called photocurrent gain, which is also often used in photodetector-device characterization. For our device, the responsivity is determined to $R_{637nm} = 0.32$ A/W for a light power density of 0.19 µW/cm² and an applied voltage of 10 V. From this the external quantum efficiency (EQE) follows to be 62%. Due to the power-law relation in equation (1), the responsivity and also the EQE decrease with increasing power density of the light.

A further important figure of merit is the specific detectivity which can be expressed as:

$$D^* = \frac{\sqrt{A \cdot B}}{NEP} \qquad (3)$$

where $A$ is again the area, $B$ is the electrical bandwidth, and NEP is the noise equivalent power. The NEP represents the minimum incident light power that a detector can distinguish from the noise. It is defined as:

$$NEP = \frac{i_n}{R_\lambda} \qquad (4)$$

where $i_n$ is the noise current of the device. Assuming a shot-noise limitation $i_{sn} = \sqrt{2eI_{dark}B}$ the specific detectivity can be written as:[43]

$$D^* = \frac{R_\lambda \sqrt{A}}{\sqrt{2eI_{dark}}} \qquad (5)$$

For our QNW photodetector this leads to a considerably high specific detectivity of $D^* > 4 \cdot 10^{13}$ Jones.

The exceptionally high on/off-ratio and specific detectivity of our QNW device is based on its very small dark current of just 1-3 pA, which is a consequence of the peculiar design of the device. Due to their large surface-to-volume ratio, the QNWs exhibit a high density of surface states that results in a Fermi-level pinning at the surface. This leads to a depletion layer and a radial band bending.[44,45] In general, the conductivity of the wires decreases with an increasing depletion layer thickness. For example, it was shown for GaN NWs grown by molecular beam epitaxy that diameters below 80 nm lead to a complete depletion of the wires, which in turn results in insulating behavior in darkness.[46] We expect that our SLS-grown CdSe QNWs with a diameter of below 12 nm are completely depleted, too. Furthermore, in a NW device that is built of many NWs organized in a network, the radial band bending leads to NW-NW tunnel-junction barriers. Finally, we expect Schottky barriers between the QNWs and both the Pt contacts and the Bi NPs. The formation of Schottky contacts between Pt and CdSe has been evidenced before in Refs. [47,48] The depleted QNWs, the QNW-QNW junctions, and the Schottky barriers at the electrodes are responsible for the low dark current of our devices. Upon illumination the photo-generated charge carriers decrease the thickness of the depletion layer and open up a conductive channel. This channel grows with increasing charge carrier generation. In addition, the increased charge-carrier density reduces the QNW-QNW junction barriers as well as the heights and the widths of the Schottky barriers.[49,50] All these effects together give rise to the increase in photocurrent of the device upon illumination.

In order to test the long term stability of the QNW photodetector device, we measured its photocurrent when biased by 10 V and while illuminated with an electrically chopped laser with a power density of 0.65 mW/cm$^2$ over a time period of 300 min. The chopper frequency was set to about 8 mHz. On and off periods were about 51 and 74 s, respectively. Figure 2d shows the measured photocurrent during the first and the last three cycles of the measurement. During each cycle fast rise and fall times for the current can be observed that will be discussed in detail in the next section. Here, we want to point to the regularity of the measured *I-t* curve. Both, the dark- and the photocurrent reveal nearly the same values for each cycle, for the first cycle to the last one after 300 min. This proves the excellent long-term stability of the QNW photodetector device under ambient conditions. This, combined with the simple power-law dependence of the photocurrent over a wide range of light intensities and a high on/off-ratio or photosensitivity of more than $10^7$ makes such a device an outstanding candidate for light detecting applications in the visible range.

**CdSe-QNW Device as a Fast Optical Switch.** The high on/off-ratio of the QNW device suggests its applicability as an optical switch. Figure 3a shows details of an *I-t* curve measured with the same parameters as the curve in Figure 2d except that the modulation frequency was increased to 0.1 Hz. The measured rise time for the current that increases from approximately $3 \cdot 10^{-6}$ µA to $4 \cdot 10^{-1}$ µA is shorter than 20 ms which is the resolution limit of the setup. The fall shows a fast and a slow component. The fast component leads to a current decrease from $4 \cdot 10^{-1}$ µA to $2 \cdot 10^{-4}$ µA within 40 ms, while it requires nearly 5 further seconds for a further decrease of the dark current to approximately $3 \cdot 10^{-6}$ µA.

We performed the same measurement using a higher laser power density of 120 mW/cm$^2$ and an increased modulation frequency of 0.25 Hz. Details of the *I-t* curve are shown in Figure 3b. The rise and the fast fall times are the same as in the previous experiment. The maximum photocurrent is increased by two orders of magnitude due to the higher illumination power density. The dark current does not reach as low values as in the previous experiment as a consequence of the shorter off period during a cycle of only 1 s. The slow component of the current fall to some extend obviously limits the on/off-ratio of the NW device for high modulation frequencies.

The physical reasons behind the drastically different components on the current fall dynamics are not unambiguously identified, yet. We suppose that the fast component is caused by the fast recombination of free charge carriers while the slowly decaying small current component is caused by the migration of trap-released charge carriers.

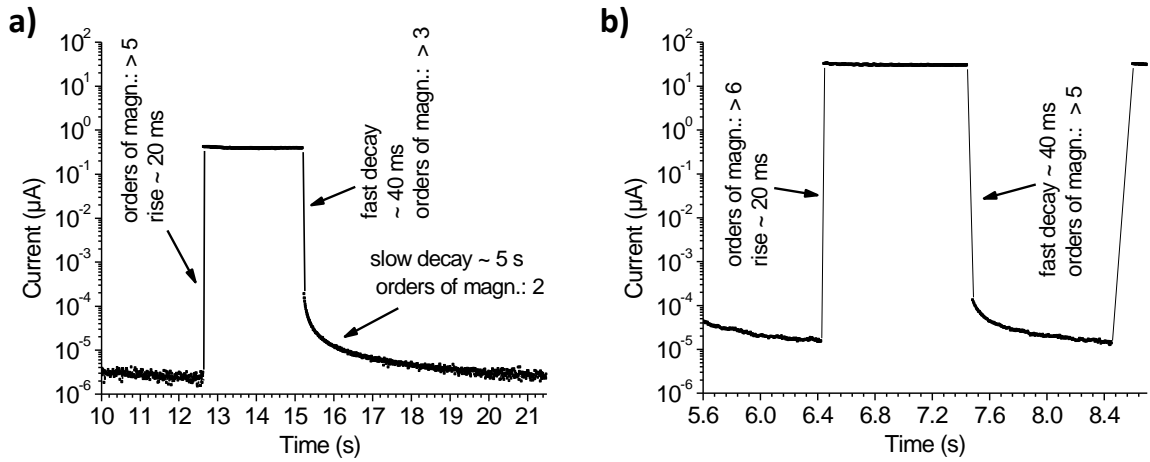

Figure 3 *I-t* curves showing rise and fall times of the device biased by 10 V and illuminated with a laser power density of (a) 0.65 mW/cm$^2$ at a asymmetric chopper frequency of 0.1 Hz and (b) 120 mW/cm$^2$ at a chopper frequency of 0.25 Hz.

The time resolution of the above experiments depends on the dynamics of the current. For moderate current changes data points are recorded every 10 ms. However, changes of the current of several orders of magnitude may decrease the resolution to about 20 ms for current jumps and 40 ms for current drops due to the integration time of the setup. In order to determine the photocurrent dynamics of our device on shorter time scales we modulated the illumination and investigated the photocurrent response by using an oscilloscope. This allows expanding the time resolution into the nanosecond regime. As a drawback, the smallest measurable currents are now about 400 nA because the current-voltage converter is not fast enough to change its sensitivity level. Note that at this point the slow part of the current-fall as shown in Figure 3 will not affect the dynamic measurements, due to its smallness in the lower pA regime and downwards.

Figure 4 shows results for the device illuminated with a maximum power density of 110 mW/cm$^2$ and a symmetric modulation frequency of up to 1 MHz. Note that the current axes are scaled linearly in these cases. Figure 4a shows the *I-t* curve for five cycles when a modulation frequency of 1 kHz is applied. Here, an almost rectangular on/off-switching behavior is observed with a maximum current of 24 µA. When the modulation frequency is increased, the rising edge starts to flatten as can be seen in Figure 4b, where the *I-t* curve for five cycles for a modulation frequency of 100 kHz is depicted. The photocurrent still reaches the maximum current during the on period of 5 µs. With further increasing modulation frequency the maximum photocurrent decreases while the minimum dark current is increasing. This leads to sinusoidal *I-t* curves as exemplarily depicted in Figure 4c for a modulation frequency of 1 MHz. Here, the maximum and minimum currents are about 19 µA and 1 µA, respectively.

Figure 4d summarizes the experiments with a chopped laser illumination. Here, the maximum photocurrent versus the chopping frequency is shown, normalized to the value of $I_{max}$ = 24 µA obtained for frequencies of up to 1 kHz. The photocurrent stays above 90% for frequencies of up to 700 kHz. A further increase of the frequency leads to a drastically decrease of the normalized photocurrent down to about 5% at 2 MHz.

A commonly used figure of merit for frequency depended measurements is the 3 dB bandwidth ($f_{3dB}$) which is defined as the frequency for which the photocurrent decreases by 3 dB to $\sqrt{1/2} \approx$ 70.7%. The 3 dB bandwidth is estimated from the diagram to be $f_{3dB} \geq 1$ MHz. This value is limited by our setup, in particular by the I/V converter which has a specific 3 dB bandwidth of 1 MHz.

With the 3 dB bandwidth, the rise time $t_{rise}$ from $10\% \cdot I_{max}$ to $90\% \cdot I_{max}$ can be calculated according to the relation:[51]

$$f_{3dB} \approx 0.35/t_{rise} \qquad (6)$$

Assuming a 3 dB bandwidth, of at least 1 MHz, the corresponding rise time is below 350 ns. Since the I-t curve is sinusoidal and symmetric, we can deduce that the rise and fall times are (roughly) the same. These fast rise and fall times can be explained by the short charge carrier lifetimes, which are in the range of few tens to hundreds ps as reported for thin SLS grown CdSe nanowires by Kuno and co-workers.[52] After the fast recombination of the carriers, the depletion layer as well as the tunneling-junctions and Schottky barriers are built-up again and the device becomes an insulator until the next light pulse creates new carriers.

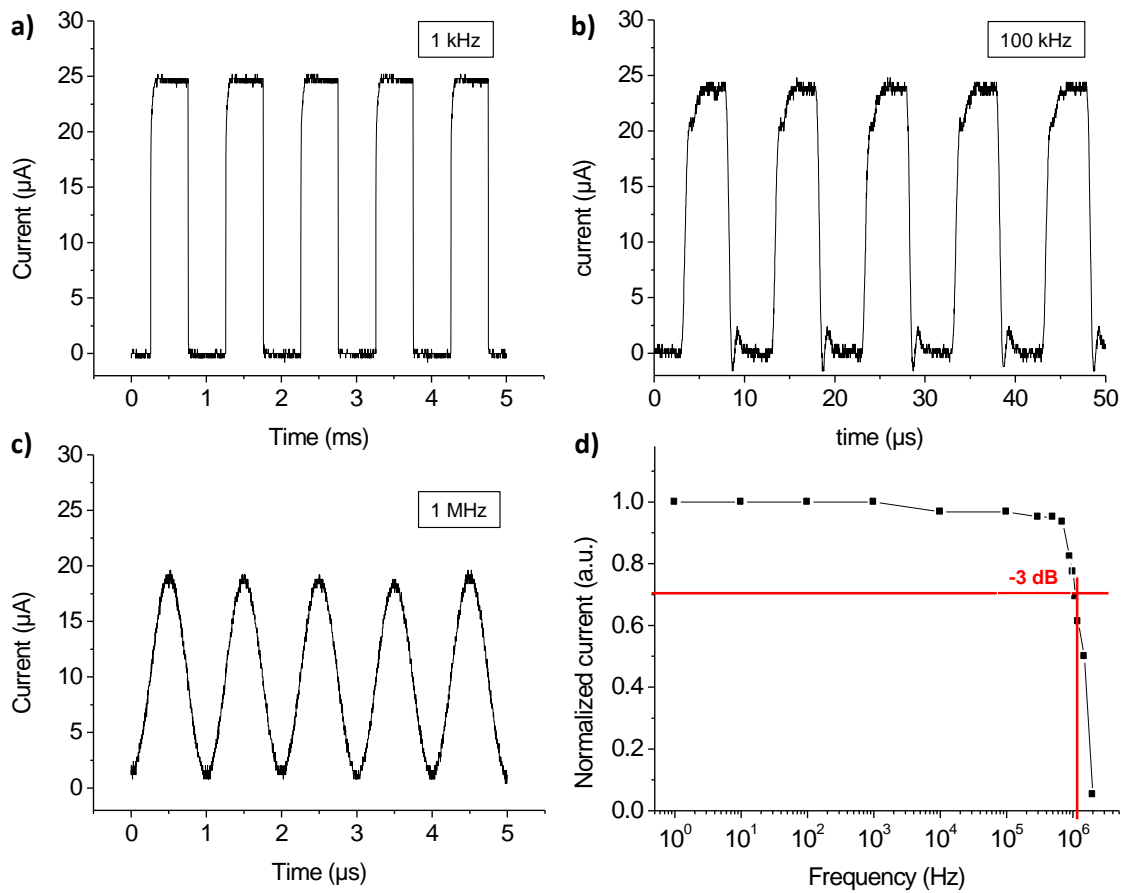

Figure 4 (a-c) Switching performance of the photocurrent for an incident laser power density of 110 mW/cm$^2$ and an applied source-drain voltage of 10 V at a chopping frequency of (a) 1 kHz Hz, (b) 100 kHz, and (c) 1 MHz. (d) Normalized current modulation amplitude versus chopping frequency. The black line connecting the data points and the red lines, which are indicating the 3 dB bandwidth and from which the rise and fall times can be deduced, are guides for the eyes.

CdSe-QNW Device Performance Compared to Other Nanostructured Photodetectors and Optical Switches. In the following we compare the performance of our CdSe-QNW device as a photodetector and as a fast optical switch to other nanostructured devices that have been reported on before. Our comparison focuses on the parameters of on/off-ratio and rise/fall times. The on/off-ratio determines the sensitivity of the photodetector, whereas fast rise/fall times are necessary to exploit a device as an optical switch. In Table 1, the on/off-ratios and corresponding illumination intensities, as well as rise and fall times (for specified on/off-ratios) for different nanostructured photodetectors based on different material compositions and morphologies are compiled. We note that the determination of rise and fall times is not uniform in literature. Very often they are defined as the time necessary for the current to rise from 10% to 90% of $I_{max}$ and to fall from 90% to 10% of $I_{max}$, respectively. These rise and fall times are also connected to the 3 dB frequency via equation (6). In Table 1, differently defined values of the rise and fall times are marked with an asterisk. For an in-depth analysis the precise definitions as given in the corresponding original references have to be regarded.

Photodetectors based on CdSe quantum dots operating in the visible range show an on/off-ratio of about $10^3$ at a laser power density of 110 mW/cm$^2$.[53] Other systems like single CdSe ribbons or CdSe-NW films exhibit on/off-ratios of $3 \cdot 10^4$ and $10^4$ at light intensities of 5 mW/cm$^2$ and 540 mW/cm$^2$, respectively.[8,54] Our CdSe-QNW photodetector exhibits an on/off-ratio of about $10^7$ under an illumination power density of 120 mW/cm$^2$.

Metal-oxide systems like single-ZnO-NW devices show exceptionally high quantum efficiencies or photocurrent gains up to $10^8$ [55–57] and also high on/off-ratios of >$10^6$ for UV light at low light intensities of below 1 mW/cm$^2$.[40] But as already discussed by Penner and co-workers[58] they reveal rise and fall times typically of several 100 of milliseconds to some seconds.[55–57,59] Fast switching systems like aligned CdSe nanocrystal-NW arrays exhibit rise and fall times of 20 µs and 30 µs.[60] However, their full on/off-ratio is only up to 100. As can be deduced from Table 1 and as has also been described in detailed comparisons by Golberg[59] and co-workers and Sargent[61] and co-workers, higher sensitivities of photodetectors are typically accompanied by a slower response, and vice versa. Our device combines both, exceptionally high on/off-ratios as well as short response times.

Table 1 Features of selected nanostructure-based photodetectors. The on/off-ratio depends on the incident power density. Rise and fall times measured between 10% and 90% of $I_{max}$, except when marked. (NW: nanowire, nc-NW: NW based on nanocrystals, NB: nanobelts, QD: quantum dots, NR: nanoribbons, n.a.: not available, *: as defined by the authors)

| Material | On/Off-ratio | P (mW/cm$^2$) | Rise (ms) | Fall (ms) | Note | Ref. |
|---|---|---|---|---|---|---|
| CdSe NW film | ~10,000 | 540 | n.a. | n.a. | | 8 |
| CdSe aligned nc-NWs array | ~ 100 | 119 | 0.02 | 0.03 | | 60 |
| P3HT:CdSe NW film | ~ 500 | 140 | <1000 | <1000 | * | 50 |
| CdSe NRod film | ~ 3 | 100 | 23 | 69 | * | 62 |
| CdSe single NR | ~18,500 | 3.41 | 0.639 | 5.68 | | 54 |
| CdSe QD film | ~ 1,000 | 110 | 0.007 | n.a. | $f_{3dB}$ = 50 kHz | 53 |
| CdS aligned NW array | ~30 | 100 | 0.8 | 240 | | 41 |
| CdS single NR | ~9,240 | 3.31 | 0.746 | 0.794 | | 51 |
| ZnS NBs | ~1,000 | n.a. | 2.57 | 2 | * | 63 |
| Zn$_3$As$_2$ aligned NW array | ~6 | 2.52 | 140 | 2900 | * | 31 |
| ZnO single NW | ~1,000,000 | 0.3 | <1000 | n.a. | * | 40 |
| β-Ga$_2$O$_3$ NW network | ~30,000 | 2 | n.a. | < 20 | * | 64 |
| Sb$_2$Se$_3$ NW film | ~150 | 14.4 | 200 | 1200 | * | 9 |
| CdSe QNW film | ~15,000,000 | 120 | < 0.00035 | < 0.00035 | $f_{3dB}$ > 1 MHz | This work |

CONCLUSION

We have demonstrated a cheap and fast fabrication route of CdSe-QNW photodetectors based on the SLS method, where QNWs were grown from Bi catalyst particles, which were prepared directly on the electrodes by electrochemical methods. This method leads to a self-limited growth of CdSe QNWs with diameters below 12 nm and lengths of more than 10 µm, with a high fraction of individual QNWs that are exposed to the light and also electrically contacted. This leads to highly sensitive photodetectors, as reflected by an on/off-ratio or a photosensitivity of more than $10^7$ over a broad range of wavelengths. The specific detectivity and responsivity are determined to $D^* = 4 \cdot 10^{13}$ Jones and $R = 0.32$ A/W, respectively.

The device can also be employed as an optical switch with a 3 dB bandwidth of more than 1 MHz, corresponding to rise and fall times below 350 ns.

The striking combination of high sensitivity and fast response is a consequence of an insulating behavior of the QNWs in the darkness and a conducting behavior due to illumination. The low dark current of approximately 3 pA at an applied bias of 10 V is caused by a combination of depleted QNWs, tunnel-junction barriers between connecting QNWs, and Schottky contacts at the electrodes. Upon illumination the radial depletion layer is reduced, the tunnel junction barriers are lowered and the QNWs become conductive. These features are consequences of the direct growth of the QNWs on the electrodes.

METHODS

**Fabrication of the CdSe QNW device.** Prior to electrode fabrication, a glass substrate was cleaned by subsequent sonication for 15 min in acetone, isopropanol, and deionized water. The finger-structured electrodes were defined by optical lithography on an area of 1 mm$^2$ (excluding the contact pads). Widths of the fingers as well as the finger interspaces were 10 µm. The electrodes were made by sputtering first 30 nm of tungsten as an adhesion layer and then 10 nm of platinum. The electro-chemical deposition of Bi NPs is analogous to our previous work.[32] In a three-electrode arrangement (HEKA model PG310 potentiostat/galvanostat) we used a platinum electrode, platinum foil, and a saturated calomel electrode as working, counter, and reference electrode, respectively. The aqueous bismuth solution with 1mM $BiCl_3$, 1 M HCl, and 50 mM KCl was purged with $N_2$ for 20 min in order to remove dissolved oxygen. By using a single pulse of -300 mV for 100 s, the Bi NPs were deposited on the electrode. CdSe QNWs were synthesized by the SLS growth method according to the process previously reported.[27] First, the 2 M selenium precursor was obtained by dissolving selenium in trioctyl phosphine (TOP) and under air-free conditions. The cadmium precursor was prepared in-situ. Therefor, CdO (26 mg, 0.2 mmol), trioctyl phosphine oxide (TOPO) (4g), and octanoic acid (OCA) (150 µL) were loaded into a four-neck flask. This mixture was dried and degassed at 100 °C under vacuum (< 1 mbar) for one hour. Then the flask was backfilled with nitrogen and heated to 280 °C to dissolve the CdO completely. Afterwards, the temperature was lowered to the reaction temperature of 200 °C. The substrate with the deposited bismuth NPs was placed into the flask and the TOP-Se (100 µL) precursor was injected. The reaction temperature was kept for 20 min before cooling down. At temperatures below 100 °C toluene was added to keep the solution liquid. After withdrawing the substrate with the grown QNWs from the flask it was immersed in toluene to purify and remove TOPO.

**Characterization of the Bi NPs and CdSe QNWs.** The bismuth NPs were imaged by an AFM (Nanowizard II from JPK Instruments). The SEM images of the CdSe QNWs were obtained on a Zeiss Evo MA10. For the photoluminescence and the Raman spectra a home-built confocal laser microscope with an $Ar^+$-laser operated at 488 nm was used.

**Characterization of the CdSe QNW photodetector.** All opto-electrical measurements were performed in a probe station (Lakeshore VFFTP4) under ambient conditions. The wavelength dependency measurement was carried out with a white-light source, a monochromator and a semiconductor characterization system (Keithley 4200-SCS). All other measurements were carried out with a diode laser (Coherent cube) operated at 637 nm. The diode laser was modulated with a TTL signal generated by the waveform generator Agilent 33120A.

The time dependent measurements were performed with an Agilent semiconductor device analyzer (B1500A), Keithley sourcemeter 2401, I/V converter (current pre-amplifier) Stanford RS570, and an oscilloscope (Tektronix TDS 2014b) with an applied source-drain voltage of 10 V. The incident laser powers were measured with an optical power meter (ThorLabs PM100A).

AUTHOR INFORMATION

E-mail: kipp@chemie.uni-hamburg.de

*Acknowledgement.* The authors thank Dino Behn for the characterization of the samples by photoluminescence and Raman spectroscopy. The work was supported by the Deutsche Forschungsgemeinschaft via Grant Nos. KI 1257/2 and ME 1380/16. Hauke Lehmann and Christian Klinke thank the European Research Council (Seventh Framework Program FP7, Project: 304980 / ERC Starting Grant 2D-SYNETRA) for their financial support. Christian Klinke acknowledges the Deutsche Forschungsgemeinschaft for a Heisenberg scholarship (KL 1453/9-1).